\documentstyle[12pt]{article}

\begin{document}

\begin{titlepage}
\begin{flushright}
OSU-HEP-00-02\\
BA-00-19\\
\end{flushright}
\vskip 2cm
\begin{center}
{\Large\bf A Mass Relation for Neutrinos}
\vskip 1cm
{\normalsize\bf
K.S.\ Babu$\,{}^1$ and S.M. Barr$\,{}^{2}$} \\
\vskip 0.5cm
{\it ${}^1\,$Department of Physics, Oklahoma State University\\
Stillwater, OK~~74078, USA\\ [0.1truecm]
${}^2\,$Bartol Research Institute, University of Delaware\\ Newark, DE 19716, USA\\[0.1truecm]
}

\end{center}
\vskip 2.5cm

\begin{abstract}

A generalization of the well-known Georgi-Jarlskog relation
$(m_{\mu}/m_{\tau}) = 3~ (m_s/m_b)$ to neutrinos is found in the
context of $SO(10)$. This new relation is $(m_{\nu_{\mu}}/m_{\nu_{\tau}})
= 16~ (m_c/m_t)$, which is consistent with present data, assuming the
small-angle MSW solution to the solar neutrino problem.

\end{abstract}

\end{titlepage}

\newpage

\noindent
{\bf The problem with $M_R$}

\vspace{0.2cm}

Very few theoretical models of fermion masses have been found in
which it is possible to predict the values or the ratios
of neutrino masses. This is in contrast to the masses of the charged
leptons and quarks, for which many fairly predictive models
exist. The problem with predicting neutrino masses lies in
the fact that they differ in character and
presumably in origin from the other fermion masses and are
therefore difficult to relate to them. This is evident
in the generally favored framework for understanding neutrino mass,
the seesaw mechanism \cite{seesaw}.

In the seesaw mechanism the neutrino mass matrix $M_{\nu}$
comes from two more basic mass matrices: a Dirac mass matrix $N$,
which couples the left-handed neutrinos to the right-handed ones,
and a Majorana mass matrix $M_R$, which couples the right-handed neutrinos
to themselves. The seesaw formula is
$M_{\nu} = - N^T M_R^{-1} N$. There are many models in which
the Dirac neutrino mass matrix $N$ is related by symmetry
to other mass matrices about which something is known,
in particular the Dirac mass matrices of the up quarks,
down quarks, and charged leptons (which we will denote by $U$, $D$, and
$L$, respectively). The problem lies with the
Majorana mass matrix $M_R$, about which most theoretical frameworks
have very little to say.
In the absence of information about
$M_R$ it is impossible to predict neutrino masses.

What is usually done is to parametrize the unknown matrix $M_R$ or
to make some ansatz for it. What is attempted here is more ambitious,
namely to find a framework in which $M_R$, as well as $N$, is
related by symmetry to the other mass matrices in such a way that a direct
and definite prediction of neutrino mass ratios becomes possible.

There is an empirical reason for expecting that $M_R$ might be related
to the Dirac mass matrices. The Dirac masses of the quarks and charged
leptons exhibit inter-family ``hierarchies", and so it seems most likely that
the Dirac neutrino mass matrix $N$ has a similar hierarchy. If there were
no hierarchy in $M_R$, then the seesaw formula would imply a hierarchy
in $M_{\nu}$ that went as the {\it square} of the Dirac mass hierarchies.
That is why it was long expected that $m_{\nu_{\mu}}/m_{\nu_{\tau}}
\sim (m_c/m_t)^2$ \cite{quad}. However, it is now known that $(m_c/m_t)^2 \sim
6 \times 10^{-6}$, whereas $m_{\nu_{\mu}}/m_{\nu_{\tau}}$ is in the range
$10^{-1}$ to $3 \times 10^{-3}$, depending on which solution (MSW oscillation or
vacuum oscillation)  of the solar
neutrino problem is correct. This suggests that the hierarchy in $M_{\nu}$
goes as the first power of the Dirac mass hierarchies rather than
quadratically. From the seesaw
formula one sees that this implies that $M_R$ has a hierarchy similar
to that of $N$ and the other Dirac matrices. This similarity in
structure suggests a direct link among all the mass matrices.

The formulas we obtain for neutrino mass ratios are indeed linear rather
than quadratic. In fact, we find that
\begin{eqnarray}
({m_{\nu_{\mu}}/m_{\nu_{\tau}}})
&=& 16 ~({m_c/m_t}) \nonumber \\
({m_{\nu_e}/m_{\nu_{\mu}}}) &=& {1 \over (16)^2}~( {m_u/m_c})~.
\end{eqnarray}
These can be thought of as natural generalizations
of the well-known Georgi-Jarlskog formulas \cite{gj}:
\begin{eqnarray}
({m_{\mu}/m_{\tau}})
&=& 3~ ({m_s/m_b}) \nonumber \\
({m_e/m_{\mu}}) &=& {1 \over (3)^2}~ ({m_d/m_s})~.
\end{eqnarray}
The reason for the
powers of 16 instead of 3 has to do with the group theory of $SO(10)$
and will be explained later.
The Georgi-Jarlskog formulas relate to each other the masses
of down-type fermions, i.e. those that are the $I_3 = -1/2$ components
of weak doublets. The formulas we obtain are the analogous relations
for the corresponding up-type fermions, i.e. those with $I_3 = +1/2$.

\vspace{0.5cm}

\noindent
{\bf The Georgi-Jarlskog formulas}

\vspace{0.2cm}

In $SU(5)$ unification the down-quark mass matrix $D$ and the charged-lepton
mass matrix $L$ are closely related to each other. In the so-called ``minimal"
scheme of Yukawa coupling they are in fact simply transposes of each other:
$L = D^T$. Thus minimal $SU(5)$ gives the predictions
$m_{\tau}^0 = m_b^0$, $m_{\mu}^0 = m_s^0$, and $m_e^0 = m_d^0$.
(The superscript zero refers here and throughout to quantities
evaluated at the unification scale.) The first of these predictions \cite{chanowitz}
is consistent with experiment, but the second and third are not.
What Georgi and Jarlskog observed is that in non-minimal
$SU(5)$ schemes some of these predictions can get modified by
group-theoretical (i.e. Clebsch) factors \cite{gj}. In particular they
showed that the relations $m_{\mu}^0 = 3 m_s^0$ and $m_e^0 = m_d^0/3$,
which are quite consistent with present data, can arise in a simple way.
Factors of $3$
emerge very naturally from the group theory of unification,
essentially because of the fact that there are three colors. It has
since been shown that the Georgi-Jarlskog
formulas can be obtained in a variety of simple and plausible ways in
different models \cite{variety}.

A simple example of $SU(5)$ Yukawa terms that lead to the Georgi-Jarlskog
formulas is the following: $h_{33} {\bf 10}_3 \overline{{\bf 5}}_3
\overline{{\bf 5}}_H + h_{22} {\bf 10}_2 \overline{{\bf 5}}_2
\overline{{\bf 45}}_H + h_{12}({\bf 10}_1 \overline{{\bf 5}}_2
+ {\bf 10}_2 \overline{{\bf 5}}_1) \overline{{\bf 5}}_H$. Here the
numerical subscripts are family indices, and the subscript $H$ refers
to a Higgs field. These terms lead to the matrices

\begin{equation}
D = \left( \begin{array}{ccc} 0 & C_d & 0 \\ C_d & B_d & 0 \\ 0 & 0 & A_d
\end{array} \right), \;\;\;\; L = \left( \begin{array}{ccc}
0 & C_d & 0 \\ C_d & -3B_d & 0 \\ 0 & 0 & A_d \end{array} \right).
\end{equation}

\noindent
The relative factor of $-3$ in the (22) elements of $L$ and $D$ comes
from a ratio of Clebsch coefficients in the coupling of the
Higgs field that is a rank-three tensor of $SU(5)$
(i.e. the $\overline{{\bf 45}}_H$).
This factor of $-3$ directly gives the prediction $m_{\mu}^0 =
3 m_s^0$. The other Georgi-Jarlskog formula arises from the fact that
$m_e^0 = -C_d^2/(-3B_d)$ and $m_d^0 = - C_d^2/B_d$.

It is to be observed, however, that $SU(5)$ relates only $L$ and $D$
to each other.
It does not relate the up-quark mass matrix $U$ or the neutrino
Dirac mass matrix $N$ to any other mass matrices; they remain quite free.
That means that if one is to extend the Georgi-Jarlskog formulas to
neutrinos a more powerful symmetry is needed. Thus we turn to $SO(10)$.

\vspace{0.5cm}

\noindent
{\bf Extending to $\boldmath{SO(10)}$}

\vspace{0.2cm}

The group $SO(10)$ relates all four Dirac mass matrices, $N$, $U$,
$D$, and $L$, to each other. In the minimal scheme of Yukawa coupling,
in which all masses come from terms of the form ${\bf 16}_i {\bf 16}_j
{\bf 10}_H$, $SO(10)$ predicts that $N = U \propto D = L$, with
the further requirement that these matrices be symmetric. This would
give the predictions $m_c^0/m_t^0 = m_s^0/m_b^0 = m_{\mu}^0/m_{\tau}^0$
and $m_u^0/m_c^0 = m_d^0/m_s^0 = m_e^0/m_{\mu}^0$.  There is too much
symmetry here and these relations are not
realistic. However, as with $SU(5)$,
group-theoretical factors different from $1$ can be introduced into some
of these relations by going to non-minimal Yukawa coupling schemes.
For example, the generalization to $SO(10)$ of the simple Yukawa operators
that gave rise to Eq. (3) is the following: $h_{33} {\bf 16}_3 {\bf 16}_3
{\bf 10}_H + h_{22} {\bf 16}_2 {\bf 16}_2 \overline{{\bf 126}}_H + h_{12}
{\bf 16}_1 {\bf 16}_2 {\bf 10}_H$.
This would lead to the same forms for $D$ and $L$
given in Eq. (3) as well as the following forms for the other Dirac
mass matrices:

\begin{equation}
U = \left( \begin{array}{ccc} 0 & C_u & 0 \\ C_u & B_u & 0 \\ 0 & 0 & A_u
\end{array} \right), \;\;\;\; N = \left( \begin{array}{ccc}
0 & C_u & 0 \\ C_u & -3B_u & 0 \\ 0 & 0 & A_u \end{array} \right),
\end{equation}

\noindent
where $C_u/A_u = C_d/A_d$. One sees that the same ratio of $-3$ between the
(22) elements of $N$ and $U$ exists as between the (22) elements of
$L$ and $D$. (This is easily understood in terms of the
$SU(4)_c \times SU(2)_L \times SU(2)_R$ decomposition of $SO(10)$,
under which the
VEV in $\overline{{\bf 126}}_H$ that
breaks the weak interactions is in the $({\bf 15}, {\bf 2}, {\bf 2})$,
and therefore couples proportionally to the $SU(4)_c$ generator $B-L$.)
For the same reasons, some of the other ways that the Georgi-Jarlskog
factor of 3 can arise in the $D$-$L$ sector also
give rise to parallel factors of 3 in the $U$-$N$ sector.
For example, if the term ${\bf 16}_2 {\bf 16}_2
\overline{{\bf 126}}_H$ is replaced by ${\bf 16}_2 {\bf 16}_2
{\bf 45}_{B-L} {\bf 45}_{I_{3R}} {\bf 10}_H$, where $\langle {\bf 45}_{B-L}
\rangle
\propto B-L$ and $\langle {\bf 45}_{I_{3R}} \rangle \propto I_{3R}$,
both $L_{22}/D_{22}$ and $N_{22}/U_{22}$ are equal to $-3$
(see the first paper in [5]).

The foregoing only shows that a relation can arise in $SO(10)$ between the
up-quark mass matrix $U$ and the Dirac mass matrix of the
neutrinos $N$, with the possibility of non-trivial group-theoretical factors.
This by itself is not sufficient to yield predictions for the
neutrinos; for that it is also necessary to be able to say something about
the Majorana mass matrix $M_R$. Happily, there does exist the possibility
in $SO(10)$ of relating $M_R$ to $U$ and $N$, as we shall now see.

Consider, for example, the Yukawa operator ${\bf 16}_i {\bf 16}_j
\overline{{\bf 126}}_H$.
The multiplet $\overline{{\bf 126}}_H$ contains both
weak-doublet components (contained in the $({\bf 15}, {\bf 2}, {\bf 2})$
of $SU(4)_c \times SU(2)_L \times SU(2)_R$), which can give the Dirac
masses for the quarks and leptons, and a weak-singlet
component (contained in the
$({\bf 10}, {\bf 1}, {\bf 3})$) that can give superlarge masses to the
right-handed neutrinos. Consequently, if both kinds of components
acquire non-vanishing vacuum expectation values, the term ${\bf 16}_i
{\bf 16}_j \overline{{\bf 126}}_H$ would contribute to all five mass
matrices $D$, $L$, $U$, $N$, and $M_R$ \cite{bm2}.
As noted before, its contributions
to $N$ and $U$ would be in the ratio $-3:1$.

Other kinds of Yukawa operators also exist that contribute both to $M_R$ and
to the Dirac mass matrices. For example, the term ${\bf 16}_i {\bf 16}_j
\overline{\bf 16}_H \overline{\bf 16}_H$ contributes to
$U$, $N$, and $M_R$ (but not to $D$ or $L$). As with
the term we previously considered, there is a Clebsch
factor between the contribution to $N$ and to $U$; however, what that
factor is depends in this case on how the fields are contracted. There are
two independent ways to contract ${\bf 16}_i {\bf 16}_j
\overline{{\bf 16}}_H \overline{{\bf 16}}_H$ to form an invariant.
The contraction that we shall consider is $[{\bf 16}_i
\overline{{\bf 16}}_H]_{45} [{\bf 16}_j \overline{{\bf 16}}_H]_{45}$,
where we mean that the fields in the brackets are contracted in the
adjoint (i.e. {\bf 45}) channel. Such a term can, of course,
arise simply by
integrating out a fermion field in the adjoint representation of
$SO(10)$. With the indices contracted in this way, it is straightforward
to show that the contributions to the elements of $N$ and
$U$ are in the ratio $3:8$.

\vspace{0.5cm}

\noindent
{\bf Obtaining mass formulas for neutrinos}

\vspace{0.2cm}

We now have the ingredients to allow a generalization of the Georgi-Jarlskog
formula to neutrinos.

First, we assume that the (22) elements of the matrices $N$ and $U$
arise from a Yukawa operator $O_{22}$ that gives them in the ratio
$-3:1$. This is motivated by the empirical success of the Georgi-Jarlskog
formulas, and by the fact already noted that some simple Yukawa terms that
give the Georgi-Jarlskog $-3:1$ ratio for the $L$-$D$ sector give the same
ratio for the $U$-$N$ sector. We also assume that the (22) element of $M_R$
remains zero. (This is automatically the case if $O_{22} =
{\bf 16}_2 {\bf 16}_2 {\bf 45}_{B-L} {\bf 45}_{I_{3R}} {\bf 10}_H$. It
is true for $O_{22} = {\bf 16}_2 {\bf 16}_2 \overline{{\bf 126}}_H$
if the weak-singlet VEV in $\overline{{\bf 126}}_H$ vanishes.)

Second, there must be enough other terms to make $M_R$ a non-singular
matrix. A particularly simple possibility would be that only the (33), (12), and
(21) elements of $M_R$ are non-vanishing. We assume that these arise from
terms of the form $[{\bf 16}_i \overline{{\bf 16}}_H]_{45} [{\bf 16}_j
\overline{{\bf 16}}_H]_{45}$,
already discussed above, with $(ij) = (12)$, and $(33)$.

With just these three terms one obtains the following realistic mass matrices

\begin{equation}
U=\left( \begin{array}{ccc} 0 & C_u & 0 \\ C_u & B_u & 0 \\ 0 & 0 & A_u \end{array}
\right), \;\; N = \left( \begin{array}{ccc} 0 & \frac{3}{8} C_u & 0 \\
\frac{3}{8} C_u & -3 B_u & 0 \\ 0 & 0 & \frac{3}{8} A_u \end{array} \right), \;\;
M_R = \left( \begin{array}{ccc} 0 & C_u & 0 \\ C_u & 0 & 0 \\ 0 & 0 & A_u
\end{array} \right) \Lambda,
\end{equation}

\noindent
where $\Lambda$ is a ratio of a GUT-scale VEV to a weak-scale VEV.
The ratios of the up quark masses can be directly read off from
the form of $U$: $m_c^0/m_t^0 = B_u/A_u$, and $m_u^0/m_c^0 \cong -C_u^2/B_u^2$.
To find the ratios of neutrino masses one must first use the seesaw
formula to compute $M_{\nu}$.

\begin{equation}
M_{\nu} = -N^T M_R^{-1} N = - \left( \begin{array}{ccc}
0 & \frac{9}{64} C_u & 0 \\ \frac{9}{64} C_u & -\frac{9}{4} B_u & 0 \\
0 & 0 & \frac{9}{64} A_u \end{array} \right) \Lambda^{-1}.
\end{equation}

\noindent
This gives the relations
$(m_{\nu_{\mu}}^0/m_{\nu_{\tau}}^0) = 16~ (m_c^0/m_t^0)$,
and $(m_{\nu_e}^0/m_{\nu_{\mu}}^0) = \frac{1}{256}~ (m_u^0/m_c^0)$,
as given in Eq. (1).
The first of these relations is obviously the more practically
interesting one, and is quite consistent with what we presently
know about neutrino oscillations. If one takes as the central value
$m_c^0/m_t^0 = 1/400$ (corresponding to $m_c(m_c) = 1.27$ GeV, $m_t^{\rm physical}=
174$ GeV, and using beta functions of minimal supersymmetry to extrapolate
the masses from $m_t$ to $M_{\rm GUT} \simeq 2 \times 10^{16}$ GeV
with $\tan\beta$ in the range of $3-30$ and $\alpha_s(m_Z) = 0.118$) \cite{pdg},
the prediction is that
$m_{\nu_{\mu}}/m_{\nu_{\tau}} \cong (25)^{-1}$. Since in this model the neutrino masses
are hierarchical, the masses of $\nu_{\mu}$ and $\nu_{\tau}$ can be
gotten directly from the $\Delta m^2$ measured in solar
neutrino oscillations and atmospheric neutrino oscillations, respectively.
For this model the small angle MSW solar solution is the relevant
one. If one takes the central value for $\Delta m^2_{21}$ to be
$5.1 \times 10^{-6}$eV$^2$ \cite{solar}, and for $\Delta m^2_{32}$ to be $3
\times 10^{-3}$eV$^2$ \cite{atmospheric}, then the central value of
the mass ratio is $m_{\nu_{\mu}}/m_{\nu_{\tau}} = (24.3)^{-1}$,
which is in remarkably good agreement with the prediction. Of course,
there are still large uncertainties in the neutrino masses, and
some uncertainty in $m_t^0$ from the running between the
unification scale and the low scale, which depends on $\tan \beta$ (this is
of order 10\% for $\tan\beta= 3-30$; the running of neutrino mass ratio
is negligible).
But presumably all of these quantities will be well measured in the
future, and the mass formula in Eq. (1) well tested.

In our framework, it is possible to predict rather precisely the masses of the heavy
$\nu_R$'s.  We find $M^R_{\nu_{\tau}} \simeq (9/64) (m_t^2/m_{\nu_\tau}) \simeq
3.7 \times 10^{13}$ GeV and $M^R_{\nu_e} \simeq M^R_{\nu_{\mu}} \simeq
(\sqrt{m_u/m_c})~ (m_c/m_t)~M^R_{\nu_{\tau}} \simeq 5.6 \times 10^9$ GeV.

How unique is the factor of 16 in Eq. (1)? Clearly,
it is no more unique than was the
factor of 3 obtained by Georgi and Jarlskog,
who could have obtained other factors --- such as 1 or 9
--- quite easily, but clearly had an eye on the actual quark
and lepton masses as well as the group-theoretical possibilities of
$SU(5)$. Similarly, by choosing different Yukawa
operators we could have gotten other factors than 16. However, we
have surveyed the possibilities and find that most other choices lead
to factors that are either much too small or much too large to be realistic.
Moreover, the choice that gives 16 also gives mass matrices that look
the most similar to the matrices of Georgi and Jarlskog.
We should mention, however, one other possibility that is also
potentially realistic. If in Eq. (5) one sets $U_{22}$ and $N_{22}$
to zero and takes instead $U_{23} = U_{32} = N_{23} = N_{32}$, as
would arise from ${\bf 16}_2 {\bf 16}_3 {\bf 10}_H$, for example,
then the relation $(m_{\nu_{\mu}}/m_{\nu_{\tau}}) \simeq (64/3) (m_c/m_t)$
results.

Another point that should be emphasized is that by relating $M_R$
and $N$ to other mass matrices, and thus predicting them precisely,
one opens up the possibility of exact predictions for the neutrino
mixing angles as well. What these predictions are will depend, of
course, on the way that the charged leptons are incorporated in the
model, so that different models that give the relation in Eq. (1)
can have different predictions for the neutrino angles.
We will now show that the charged leptons and down quarks can be
accommodated in a simple fashion in the present framework.

\vspace{0.5cm}

\newpage

\noindent
{\bf Including the charged leptons and down quarks}

\vspace{0.2cm}

There is more than one way to include the charged leptons and down
quarks in the present framework. Here we present an example that
gives realistic predictions, is fairly simple, and incorporates
the Georgi-Jarlskog formulas. The structure is quite similar
to the model proposed in \cite{ab2}. The matrices $U$, $N$, and
$M_R$ are as given in Eq. (5), the remaining mass matrices are
(in a notation where the right--handed singlet fermions $f^c$ multiply
on the left and the left--handed doublet fermions $f$ on the right)
\begin{equation}
D = \left( \begin{array}{ccc} 0 & C_d & 0 \\
0 & B_d & A^{\prime}_d \\ C'_d & 0 & A_d \end{array}
\right), \;\;\;\;  L = \left( \begin{array}{ccc} 0 & 0 & C'_d \\
C_d & -3 B_d & 0 \\ 0 & A'_d & A_d \end{array}
\right).
\end{equation}

\noindent
The (22) elements come from the same term as gives the (22) elements of
$U$ and $N$, which can be, for instance, ${\bf 16}_2 {\bf 16}_2
\overline{\bf 126}_H$ or ${\bf 16}_2 {\bf 16}_2 {\bf 45}_{B-L}
{\bf 45}_{I_{3R}} {\bf 10}_H$. The ``lopsided" (23)[(32)]
elements in $D[L]$ come
from $[{\bf 16}_2 {\bf 16}_H]_{10} [{\bf 16}_3 {\bf 16}'_H]_{10}$, in
a notation explained previously.
Here the ${\bf 16}_H$ acquires a GUT scale
VEV along the weak singlet direction, while the ${\bf 16'}_H$ has a VEV
along the weak doublet.
Such lopsided elements have been
shown to explain in a simple way the largeness of the atmospheric
neutrino mixing and the smallness of $V_{cb}$ \cite{lopsided}.
Similar terms generate the other lopsided
off--diagonal entries of $D,L$.

The hierarchy in masses follows if we assume $C_d,C_d' \ll
B_d \ll A_d,A_d'$.  Denote the ratios $A_d'/A_d \equiv
\sigma,~ B_d/A_d \equiv \epsilon/3,~ C_d/A_d \equiv \delta,$
and $C_d'/A_d \equiv \delta'$ with $(\delta \sim \delta')
\ll \epsilon \ll (\sigma \sim 1)$.
After untangling the large $(s_R-b_R)$ mixing in $D$
and the $\mu_L-\tau_L$ mixing in $L$
(both of which are parameterized by a common angle
$\tan\theta \equiv \sigma$), the matrices in Eq. (7) will look
very much like those in Eq. (3).  All phases in $D$ and $L$ can be
removed by field redefinitions; a single phase $\phi$ will remain in
$U$ and an independent phase $\phi'$ in $M_\nu$.
(We take these phases to be in the (12) and (21) elements.)
For the mass ratios in the down sector we obtain:
$b/\tau \simeq 1$, $s/\mu \simeq 1/3$, $d/e
\simeq 9$,
$\mu/\tau \simeq \epsilon/(1+
\sigma^2)$, and $e/\mu \simeq \delta
\delta'\sigma\sqrt{1+\sigma^2}/\epsilon^2$.
(Here and below we denote the mass of a particle at the GUT scale by
the particle's name, eg. $s \equiv m_s^0$.)
The first of these relations
is the well-known successful mass relation of minimal $SU(5)$
\cite{chanowitz}, while
the second and third are the Georgi-Jarlskog relations \cite{gj}.

For the quark mixing angles we obtain:
$|V_{us}^0| \simeq \left|\sqrt{d/s}\sqrt{\sigma} (1+\sigma^2)^{-1/4}
xe^{i\phi} + \sqrt{u/c} \right|$,
$|V_{cb}^0| \simeq (s/b) \sigma$, and
$|V_{ub}^0| \simeq (s/b)\left|\sqrt{d/s}
\sigma^{-1/2}(1+\sigma^2)^{-1/4} x e^{i \phi}
- \sqrt{u/c} \sigma\right|$, where $x \equiv \sqrt{\delta'/\delta}$.
From these the values of the model parameters can be extracted,
allowing then a prediction
of the CP-violating Wolfenstein parameter $\eta$, using the following
relation:

\begin{equation}
\eta^0 \equiv {\rm Im}\left({V_{ub}^0 V_{cs}^0 \over V_{us}^0 V_{cb}^0}\right)
\simeq
-{1 \over |V_{us}^0|^2} \sqrt{u/c} \sqrt{d/s}
\sigma^{-3/2} (1+ \sigma^2)^{3/4} x \sin\phi~.
\end{equation}

The parameter $\sigma$ and $\epsilon$ are best determined from
$3|V_{cb}^0|/(\mu/\tau)$
and $\mu/\tau$ respectively.  Besides the input value of $|V_{cb}|$ at the weak
scale, we need the renormalization factor to extrapolate it to $M_{\rm GUT}$.  For
illustration we choose $|V_{cb}| = 0.036 $ and the RGE factor to be $0.885$, corresponding
to $\tan\beta = 3-30$ in MSSM. ($|V_{ub}/V_{cb}|, ~\eta, ~|V_{us}|$,
and $m_\mu/m_\tau$ are approximately RGE invariant).
Then $\sigma \simeq 1.6$ and $\epsilon\simeq 0.21$.  This explains
naturally why the $\mu-\tau$ mixing angle is large ($\tan\theta = \sigma$) while
$V_{cb}$ is small ($V_{cb} \sim s/b$).
If we fit $|V_{us}^0| = 0.22$ and $|V_{ub}^0| = 0.0025$ (corresponding to a central
value of $|V_{ub}/V_{cb}|= 0.08$), we find $x\cos\phi = -0.64$ and $x = 1.27$.
The Wolfenstein parameter $\eta$ is then predicted to be $\eta = 0.36$, which
is right at its central value \cite{pdg}.  This non--trivial relation can be considered
as another success of the model.

The fact that $M_R$ has been nailed down, except for the phase $\phi'$,
means that there are
predictions for the mixing angles of the neutrinos as well as for their mass
ratios.  We obtain:
$|U_{\mu 3}| \simeq \sigma/\sqrt{1+\sigma^2} \simeq 0.85$,
$|U_{\mu 2}| \simeq 1/\sqrt{1+\sigma^2} \simeq  0.53$,
$|U_{e2}| \simeq \left|\sqrt{e/\mu}
\sigma^{-1/2}(1+\sigma^2)^{-1/4} x^{-1}e^{i \phi'} - \sqrt{\nu_e/\nu_{\mu}}
\right|  = 0.031 \pm 0.004$, and
$|U_{e3}| \simeq |\sqrt{e/\mu} \sigma^{1/2}$ $(1+\sigma^2)^{-1/4}
x^{-1}| \simeq 0.05$. For the leptonic phase
$\eta_\ell \equiv {\rm Im}\left({U_{\tau 1} U_{\mu 2} \over U_{\mu 1}
U_{\tau 2}}\right)$ one has $\eta_{\ell} \simeq {\rm Im}
(1 - $ $\sqrt{e/
\mu}\sqrt{\nu_{\mu}/ \nu_e} \sigma^{-1/2}(1+\sigma^2)^{3/4} x^{-1}
e^{i \phi'} )^{-1}$, which is bounded to be less than $3 \times
10^{-2}$ in absolute value.

The solar neutrino oscillation is governed by $4 |U_{e2}|^2(1-|U_{e2}|^2)
\simeq 3.8 \times 10^{-3}$.  This value is nicely consistent with the small angle
MSW explanation the solar data \cite{solar}.  We note that in this model leptonic
CP violation is relatively small.  As for the atmospheric oscillation angle, we obtain
$\sin^22\theta_{\rm atm}^{\rm osc} \simeq 0.81$.  While this value is in the right
range, it is somewhat on the lower side in this illustrative example.  If we allow
for a $15\%$ reduction in $|V_{cb}|$ due to finite chargino corrections \cite{raby} (this is
natural if $\tan\beta$ is moderately large; corrections of this order is also
suggested by $b-\tau$ unification \cite{hrs}), then $\sigma \simeq
1.3$, which will change the atmospheric oscillation angle to about $0.93$.

While the specific way we adopted for including down quarks and charged
leptons is by no means
unique, we have illustrated how it becomes possible in our framework to calculate precisely
the leptonic mixing parameters simultaneously with the neutrino mass ratios.

\vspace{0.2cm}

This work is supported in part by the Department of Energy Grant No. DE FG02 01ER 4062
and by funds from the Oklahoma State University.

\end{document}